\documentclass[letterpaper]{article}

\usepackage{prepr}

\title{Extending Decision Maps for Sustainable Safety and Security in Self-Adaptive Systems}
\author{
\PREPauthor{Marco Stadler}{Johannes Kepler University Linz, LIT Secure and Correct Systems Lab / Institute of Business Informatics -- Software Engineering}{marco.stadler@jku.at}
\PREPauthor{Wesley	K.G. Assunção}{North Carolina State University, Department of Computer Science}{wguezas@ncsu.edu}
\PREPauthor{Michael Vierhauser}{University of Innsbruck, Department of Computer Science}{michael.vierhauser@uibk.ac.at}
\PREPauthor{Iris Groher}{Johannes Kepler University Linz, Institute of Business Informatics -- Software Engineering}{iris.groher@jku.at}
\PREPauthor{Michael Riegler}{ENGEL Austria GmbH, Information Security}{michael.riegler@engel.at}
\PREPauthor{Johannes Sametinger}{Johannes Kepler University Linz, LIT Secure and Correct Systems Lab / Institute of Business Informatics -- Software Engineering}{johannes.sametinger@jku.at}
\vspace{0.6cm}
}
\setvenue{the \vspace{3pt} 7th IEEE International Conference on Autonomic Computing and Self-Organizing Systems (ACSOS)}
\setvenueone{the 7th IEEE International Conference on Autonomic Computing and Self-Organizing Systems (ACSOS)}
\setyear{2026}
\setdoi{TBA}

\usepackage{graphicx}
\usepackage{textcomp}
\usepackage{xcolor}
\usepackage{tcolorbox}

\usepackage{balance}

\usepackage{graphicx}
\usepackage{xspace}
\usepackage[scaled=.92]{helvet}
\usepackage[T1]{fontenc}
\usepackage[scaled=.92]{helvet}
\usepackage[T1]{fontenc}
\usepackage{booktabs} 
\usepackage{url}
\usepackage{multirow}
\usepackage{rotating}
\usepackage{array}
\usepackage{verbatim}
\usepackage{multirow}
\usepackage{setspace}
\usepackage{tabularx}
\usepackage{url}
\usepackage{pifont}
\usepackage{graphicx}
\usepackage{subcaption}
\usepackage[font=footnotesize]{caption}
\usepackage{array,booktabs}
\usepackage{soul}
\soulregister{\ref}{1}
\soulregister{\cite}{1}
\soulregister{\pageref}{1}
\usepackage{vcell}
\usepackage{pifont}
\usepackage{afterpage}
\usepackage{rotating}
\usepackage{wrapfig}
\usepackage{enumitem}
\usepackage{lipsum}
\usepackage{tikz}
\usepackage{listings}
\usepackage{shadowtext}
\usepackage{etoolbox}
\usepackage{harveyballs}
\usepackage{orcidlink}
\usepackage{fontawesome}
\usepackage{circledsteps}
\usepackage{float}
\usepackage[export]{adjustbox}
\usepackage{hyperref}

\newcommand{\fwcomp}[1]{\ttformat{#1\xspace}}

\newcommand{\indpart}[0]{ENGEL Austria GmbH\xspace}
\newcommand{\machines}{{injection molding machines}\xspace} 

\newcommand{\ms}[1]{{\color{blue}{#1}}}

\newcommand{\indcomp}[1]{{\setlength{\fboxsep}{2pt}\colorbox{dim-ind-color}{\fwcomp{\uppercase{#1}}}}}

\renewcommand{\indcomp}[1]{%
  {%
    \setlength{\fboxsep}{1pt}%
    \setlength{\fboxrule}{1.25pt}
    \fcolorbox{dim-ind-color}{white}{\fwcomp{\uppercase{#1}}}%
  }%
}

\definecolor{dim-ind-color}{HTML}{44AA99}

\definecolor{myblue}{HTML}{007FFF}
\definecolor{myorange}{HTML}{FF8000}
\definecolor{oliveish}{HTML}{ccbb44}
\definecolor{pinkish}{HTML}{ee6677}
\definecolor{greyish}{HTML}{bbbbbb}
\definecolor{blueish}{HTML}{66ccee}
\definecolor{greenish}{HTML}{6ab797}

\newcommand{\ttformat}[1]{{\texttt{\small #1}}}

\definecolor{alizarin}{rgb}{0.82, 0.1, 0.26}

\newcommand{\ucitem}[1]{
\noindent$\bullet$ \textbf{UC-#1 -- }
\ifthenelse{\equal{#1}{1}}{\textbf{Edge Computing:}\xspace}{\textbf{Multi-Modal Auth. System:}\xspace}%
}

\colorlet{punct}{red!60!black}
\definecolor{background}{HTML}{EEEEEE}
\definecolor{delim}{RGB}{20,105,176}
\colorlet{numb}{magenta!60!black}
\lstdefinelanguage{json}{
    basicstyle=\ttfamily\footnotesize,
    numbers=left,
    numberstyle=\scriptsize,
    stepnumber=1,
    numbersep=8pt,
    showstringspaces=false,
    breaklines=true,
    frame=lines,
    literate=
     *{0}{{{\color{numb}0}}}{1}
      {1}{{{\color{numb}1}}}{1}
      {2}{{{\color{numb}2}}}{1}
      {4}{{{\color{numb}4}}}{1}
      {5}{{{\color{numb}5}}}{1}
      {6}{{{\color{numb}6}}}{1}
      {7}{{{\color{numb}7}}}{1}
      {8}{{{\color{numb}8}}}{1}
      {9}{{{\color{numb}9}}}{1}
      {:}{{{\color{punct}{:}}}}{1}
      {,}{{{\color{punct}{,}}}}{1}
      {\{}{{{\color{delim}{\{}}}}{1}
      {\}}{{{\color{delim}{\}}}}}{1}
      {[}{{{\color{delim}{[}}}}{1}
      {true}{{{\color{numb}{true}}}}{1}
      {]}{{{\color{delim}{]}}}}{1},
}

\newcounter{rqcounter}

\makeatletter
\patchcmd{\thebibliography}
  {\section*{\refname}}%
  {\section*{\refname}}
  {}{}
\makeatother

\definecolor{reviewbg}{RGB}{188, 222, 255} 
\sethlcolor{reviewbg}

\newcommand{\hball}[1]{\raisebox{-0.15em}{\csname harveyBall#1\endcsname}}

\DeclareCaptionLabelFormat{purplelabel}{\textcolor{purple}{#1~#2:}}

\newtcbox{\inlinehlbox}{on line,
  colback=reviewbg, colframe=reviewbg,
  boxrule=0pt, boxsep=0pt,
  left=0pt, right=0pt, top=0pt, bottom=0.75pt,
  enhanced, sharp corners}

  \newtcolorbox{highlighted}{enhanced,
  breakable,
  sharp corners,
  colback=reviewbg,
  colframe=reviewbg,
  boxrule=0pt,
  boxsep=0pt,
  left=1pt,
  right=1pt,
  top=0pt,
  bottom=0pt}

\tcbset{
  myrqbox/.style={
    boxrule=1pt,
    arc=6pt,
    left=2mm,
    right=2mm,
    top=1mm,
    bottom=1mm,
    fonttitle=\bfseries,
  }
}

\begin{document}

\maketitle

\begin{abstract}
    Sustainability refers to a system's ability to maintain its functionality and endure over time. Hence, sustainability is a highly desirable property of software systems, including Self-Adaptive Systems (SASs). SASs can change (adapt) their behavior at runtime to continue achieving their objectives despite external or internal impacts.
    SASs' intended long-term system behavior can be expressed through a sustainability-driven visual modeling notation called Decision Maps (DMs).
    Although DMs have been proven helpful, they lack adequate modeling support for safety and security concerns.
    We address this limitation by extending the current notation for sustainability-driven modeling of SASs to better accommodate the unique characteristics of safety and security scenarios. 
    First, we introduce an additional modeling dimension to account for safety incidents.
    Second, we adopt a fine-grained divide-and-conquer approach, modeling from distinct temporal security viewpoints (``security modes'') to address security.
    We employ the extended DM notation in a real-world use case scenario provided by our industry partner to assess its feasibility and suitability for practitioners.
    Our results indicate that our modeling notation helps capture security and safety scenarios more accurately and provides holistic support for the self-adaptation life cycle phases.
    
\end{abstract}


\section{Introduction}

Software systems that handle distributed applications in dynamic environments typically require human oversight to operate both reliably and safely~\cite{salehieSelfadaptiveSoftwareLandscape2009}. 
Self-adaptive systems (SASs) commonly address this challenge by employing a feedback loop mechanism that autonomously adjusts to, for example, environmental changes to preserve the system's utility without constant human intervention~\cite{salehieSelfadaptiveSoftwareLandscape2009}.
In many domains, SASs often also exhibit safety- and security-critical characteristics~\cite{pekaricSystematicReviewSecurity2023,weissIntegratingUndependableSelfadaptive2018,chehidaModelbasedApproachSelfadaptive2024}.
For instance, a self-adaptive robotic system should never harm the humans working in close proximity (safety), nor should it be possible for a malicious actor to trigger such behavior (security).
These characteristics can be denoted as software \textit{intents}, describing the essential sustainability boundaries on, and expectations of a system's behavior~\cite{huismanSoftwareThatMeets2016,lagoNewVisionSoftware2026}.
Hence, for a software system to preserve safety and security over time, it must meet its corresponding security and safety intents.

While SAS engineering addresses primarily long-term technical sustainability (e.g., maintainability~\cite{ventersSoftwareSustainabilityModern2014}), runtime adaptation typically optimizes for short-term technical properties~\cite{gerostathopoulos2022} such as performance, reliability, or availability~\cite{ weynsPatternsDecentralizedControl2013, wongSelfadaptiveSystemsSystematic2022,donakantiReimaginingSelfAdaptationAge2024}.
Such a short-term focus often neglects a holistic sustainability perspective that explicitly balances safety and security as core pillars alongside environmental and social concerns.

Sustainability modeling~\cite{lagoSustainabilityAssessmentFramework2025} offers a complementary perspective, framing adaptation as a \textit{long-term}, \textit{value-driven process}, rather than an immediate technical reaction.
In this context, \textit {sustainability} refers to a system’s ability to endure and maintain effective operation over prolonged periods~\cite{gerostathopoulos2022,lagoSustainabilityAssessmentFramework2025,vermeulenConnectedWorldInsights2023,lagoNewVisionSoftware2026}. 
Anderson~\cite{andersonMakingSecuritySustainable2018} defines sustainability in the context of security as the capability to maintain safety-critical security properties over the long operational lifetime of systems through continuous update, maintenance, and governance.

The combination of sustainability and self-adaptation allows for modeling the adaptation intent as a sustainability goal aimed at achieving the long-term success of an SAS~\cite{gerostathopoulos2022}. 
The core idea hereby is to express the adaptation intent at design time using a visual notation called Decision Maps (DMs)~\cite{lagoArchitectureDesignDecision2019} to analyze its impact both at runtime and over prolonged periods of time~\cite{gerostathopoulos2022}.

Continuously monitoring a system and its environment, analyzing the monitored data, and determining an adaptation action are among the significant challenges faced by practitioners~\cite{gerostathopoulos2022}. 
This problem becomes even more complex in safety- and security-critical environments~\cite{pekaricSystematicReviewSecurity2023}, requiring guidance and proper tool support to assess and integrate safety and security impacts.
Current DM-based modeling techniques~\cite{lagoSustainabilityAssessmentFramework2025,gerostathopoulos2022} capture adaptation intents, but cannot express fine-grained sustainable safety or security concerns. 
Furthermore, security incidents cause a shift in priorities at runtime, which is not covered by existing sustainability models~\cite{lagoSustainabilityAssessmentFramework2025}.
Thus, SASs operating in safety- and security-critical environments need extended modeling capabilities that combine sustainability goals with system states during an attack. 
Therefore, our work is motivated by the following research questions (RQs):

\begin{itemize}[leftmargin=*]
    \item \textbf{RQ1:} \emph{What concepts and extensions are needed to enhance the sustainability-driven DM modeling notation so that it can represent safety- and security-aware adaptation intents in SASs?}
    \item \textbf{RQ2:} \emph{To what extent is the extended DM notation feasible for capturing sustainable safety- and security-relevant adaptation intents in real-world and benchmark SASs?}
\end{itemize}
    
We build on the DM modeling notation and augment it with new semantic elements to more accurately capture and sustain the long-term adaptation intent.
Our results demonstrate that the extended notation captures safety and security intents, explicitly exposing dynamic trade-offs for compensatory adaptation that overcome the limitations of static modeling.
With this work, we contribute to the broader notion of achieving long-lasting (sustainable) adaptive software systems used in both safety- and security-relevant environments.\vspace{0.2em}

\section{Background}
\label{sec:background}

  

SASs are systems capable of modifying their structure or behavior at runtime to maintain their goals despite environmental or internal uncertainties~\cite{salehieSelfadaptiveSoftwareLandscape2009}.
SASs are usually divided into a \textit{Managed system} (the application logic subject to change) and a \textit{Managing system} (the adaptation logic that performs the runtime changes)~\cite{weynsPatternsDecentralizedControl2013,arcainiModelingAnalyzingMAPEK2015,gerostathopoulos2022}, acting as a feedback loop, most commonly realized through the MAPE-K reference model, which \underline{M}onitors, \underline{A}nalyzes, \underline{P}lans, and \underline{E}xecutes adaptations over a shared \underline{K}nowledge base~\cite{weynsPatternsDecentralizedControl2013,brunEngineeringSelfAdaptiveSystems2009,kephartVisionAutonomicComputing2003}.

\textbf{Sustainability Modeling and Evolution of Quality Concerns:}
Software sustainability is defined as ``the preservation of the long-term and beneficial use of software, and its appropriate evolution, in a context that continuously changes''~\cite{lagoNewVisionSoftware2026,vermeulenConnectedWorldInsights2023}. 
Lago~\etal~\cite{lagoSustainabilityAssessmentFramework2025} developed the Sustainability Assessment Framework (SAF) Toolkit that provides support for modeling sustainability as a software quality property and includes DMs, which show sustainability trade-offs and dependencies among quality attributes.
It further contains a ``Sustainability-Quality model'', which categorizes and measures quality attributes across four dimensions:
1) \textit{social}, the integration of systems within communities taking into account their impact on society; 2) \textit{technical}, the evolution, maintenance, and long-term utilization of software systems; 3) \textit{environmental}, the impacts on the natural ecosystem, greenhouse gas emissions, etc.; and 4) \textit{economic}, the business considerations~\cite{lagoFramingSustainabilityProperty2015,lagoArchitectureDesignDecision2019,gerostathopoulos2022, lagoSustainabilityAssessmentFramework2025,lagoNewVisionSoftware2026}. 
DMs are used to illustrate the features of a selected software project that should be sustainability aware.
Stakeholders use a DM to reason about the implications of the design decisions made and their effect on the quality attributes, which are categorized using the four previously outlined dimensions. 
Effects on quality attributes can be either \fwcomp{positive}, \fwcomp{negative}, or \fwcomp{undecided}.
The quality attributes are further detailed by mapping them across the sustainability-time dimensions~\cite{hiltyICTSustainabilityEmerging2015}.
The impacts of information and communication technology (ICT) can take either direct (i.e., \fwcomp{IMMEDIATE}) effects, \fwcomp{ENABLING} effects, or \fwcomp{SYSTEMIC} effects.

    
    

\begin{figure}
    \centering
    \includegraphics[width=\linewidth]{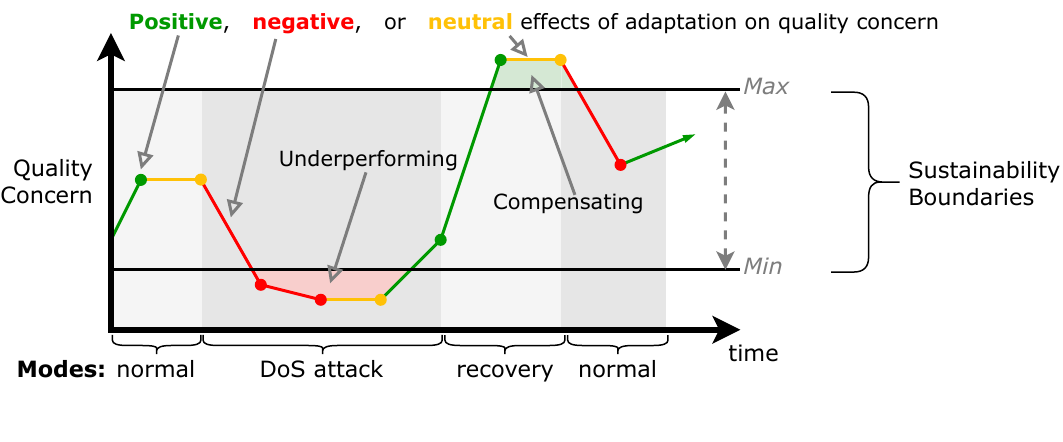}
    \vspace{-2em}
    \caption{Evolution of quality concerns over time with compensatory adaptation (adapted from~\cite{gerostathopoulos2022,lagoSustainabilityAssessmentFramework2025}).}
    \label{fig:our-intent}
    \vspace{-1em}
\end{figure}


Gerostathopoulos~\etal~\cite{gerostathopoulos2022} proposed to combine sustainability modeling with SASs, modeling the so-called adaptation \textit{intent} of a managed system as a sustainability goal.
By preserving the original intent of the managed system, the SAS can accommodate changes over time, and only then, they can be considered truly successful.
The novel idea is that an adaptation (i.e., one MAPE-K loop) in an SAS is usually only concerned with a short period of time.
In essence, adaptation can enrich its short-term perspective with a long-term perspective through sustainability~\cite{gerostathopoulos2022}.


\textbf{Security:}
Security in software systems is defined as the protection of information and resources to ensure confidentiality, integrity, and availability~\cite{nielesIntroductionInformationSecurity2017}.
Maintaining security is challenging, as it is not a static property but a dynamic state that must be actively defended.
A system may be considered ``secure'' at one point in time, and after a new vulnerability is disclosed, the attack surface can change completely.
The Log4Shell vulnerability~\cite{graham-cummingLog4j2VulnerabilityCVE2021442282021} is a prominent example, allowing remote code execution in the widely used Log4j Java logging framework: disclosed on 9~December 2021, it rendered systems that were considered secure the day before insecure overnight.
Hence, systems must co-evolve (adapt) with the threat landscape to effectively protect their resources.


SASs capable of protecting and mitigating security threats at runtime are called \textit{self-protecting software systems}~\cite{yuanSystematicSurveySelfProtecting2014}. 
One concept SASs use is \textit{security modes}, which decompose a system into multiple operational modes~\cite{rieglerDistributedMAPEKFramework2023,beggsHumancentredSecurityNew2025,romagnoliRuntimeSystemSupport2023,sisodiaSecuringSmartHome2018}.
Each mode denotes a distinct operational state defined by its own set of resource configurations and tasks~\cite{raoFIREFinelyIntegrated2022}.
For instance, a web server can be split into a \fwcomp{normal mode} and a \fwcomp{vulnerable mode}: the web server enters the \fwcomp{vulnerable mode} when a new vulnerability (like Log4Shell) is published but no fix is yet available, trading off properties such as functionality, availability, or energy efficiency for security (e.g., restricting access or up-scaling resources).
The web server switches back to \fwcomp{normal mode} once the fix is installed.
Security modes were proposed as early as 2001~\cite{goseva-popstojanovaCharacterizingIntrusionTolerant2001} and have since been used in the SASs community~\cite{ahmadAdaptiveSecuritySelfProtection2023,rieglerDistributedMAPEKFramework2023}.


\textbf{Safety:}
Safety refers to ``the absence of catastrophic consequences on the user(s) and the environment''~\cite{avizienisBasicConceptsTaxonomy2004}.
Safety and security are interconnected concepts.
Security influences the functional safety of  systems~\cite{pekaricSystematicReviewSecurity2023}.
For instance, a security issue becomes a safety risk if hackers manage to access pacemakers and alter the clock rate or set off emergency shocks, harming patients~\cite{rieglerContextAwareSecurityModes2022}.
\begin{figure}
    \centering
    \includegraphics[width=\linewidth]{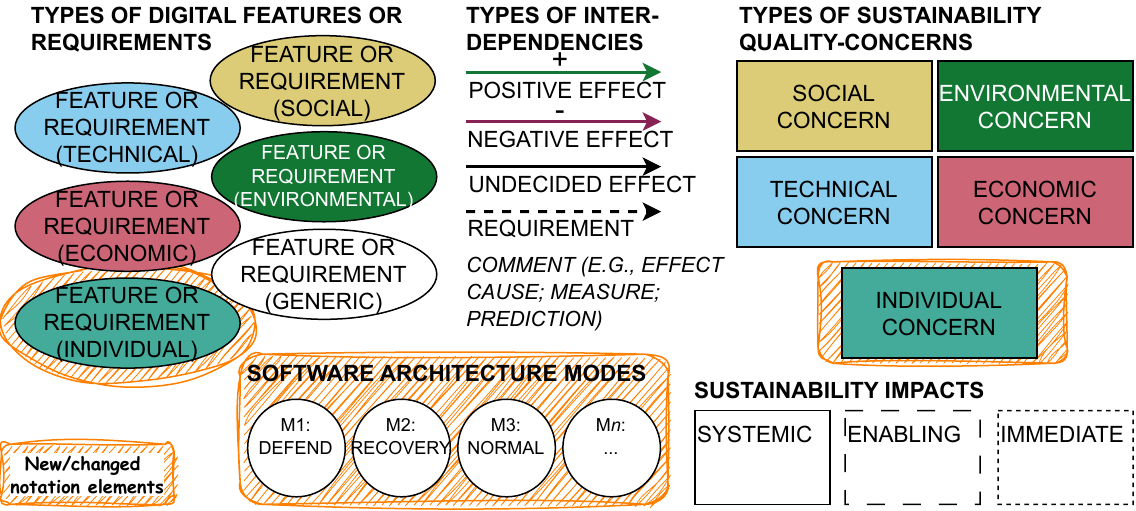}
    \caption{Extending the DM notation of Lago~\etal~\cite{lagoSustainabilityAssessmentFramework2025} (colors have been adapted for colorblind accessibility~\cite{wongPointsViewColor2011}).}
    \label{fig:our-dm-notation}
    \vspace{-1em}
\end{figure}
Penzenstadler~\etal~\cite{penzenstadlerSafetySecurityNow2014} argue that safety and security are not merely operational constraints but foundational pillars that enable a system's long-term endurance.
For systems involving humans in the loop, sustainable safety means preserving the \textit{individual's} well-being over time. 
For instance, by preventing operator fatigue, chronic stress, or cognitive overload that could lead to future accidents.
The recent ISO/IEC 25010:2023 standard~\cite{internationalorganizationforstandardizationSystemsSoftwareEngineering2023} explicitly elevates Safety to a primary quality characteristic, yet current sustainability modeling approaches~\cite{lagoSustainabilityAssessmentFramework2025} still aggregate safety and related human concerns under a broad ``social'' dimension. This aggregation limits the ability to reason about and operationalize these concerns in SASs.
Hence, SASs require a more fine-grained perspective that explicitly distinguishes and represents such concerns. This explicit representation is a prerequisite for achieving sustainable safety.




\textbf{Running Example:}
\textit{Safety} -- Machine workers are working in close proximity to a robotic system to assemble parts (``cobots'')~\cite{javaidSignificantApplicationsCobots2022}, with a supervising human-in-the-loop part of the SAS~\cite{liGenerativeAISelfAdaptive2024}.  
A human can shut down a robot if a sensor alert is detected, but they need to determine if the alert is a true positive and confirm a real emergency situation.
Although the system is designed to provide technical resilience to ensure safe operation, a fatigued operator with manual override privileges may still circumvent these safeguards.
The current dimensions of the modeling notation cannot capture this critical sustainability concern:
The \textit{social} dimension focuses on communities/collective welfare~\cite{lagoFramingSustainabilityProperty2015}, rather than individual cognitive or behavioral factors such as fatigue or situational awareness.
The \textit{technical} and \textit{economic} dimensions represent system and cost optimization; they model performance, maintainability, and resource use, but not human reliability.
The \textit{environmental} dimension focuses on ecological effects~\cite{lagoFramingSustainabilityProperty2015}.
Therefore, a person-centric perspective is necessary; sustaining this resilience over the long term requires modeling human cognitive degradation.

\textit{Security} -- Continuing with the above example, we assume that cobots are working in a warehouse production line orchestrated by an industrial control system (ICS).
For instance, consider a Denial-of-Service (DoS) attack targeting the ICS. 
A DoS attack's goal is to tamper with the availability of a target system by generating a large volume of traffic, depleting the target's infrastructure~\cite{chagantiComprehensiveReviewDenial2022}.
In a standard operational context, the sustainability goal is to minimize energy consumption to reduce the carbon footprint; therefore, a static DM would model a ``high energy'' state as a negative impact on the environmental dimension.
However, during a DoS attack, the immediate priority shifts to system resilience, with a technical focus on guaranteeing safety and availability to ensure correct operation under severe stress. 

The existing DM notation only supports modeling adaptation intents statically along four sustainability dimensions. 
Security incidents are modeled as generic disruptions, failing to capture the changing priorities or trade-offs in each attack life cycle phase.
Furthermore, environmental aspects such as energy spikes during countermeasures and normalization in recovery are indistinct, and the fact that security adapts over time, shifting from containment to restoration, is absent.

\section{Related Work}
\label{sec:relwork}

Sustainability in software engineering (SE) has gained increasing importance~\cite{matathammal2025edgemlbalancer,lagoNewVisionSoftware2026} and has been discussed in domains such as requirements engineering (RE)~\cite{bambazek2023RE}, software architecture~\cite{Fatima2024}, and development~\cite{Oyedeji2025}.
Many sustainability approaches in SE focus on assessing the potential sustainability impacts of software systems during RE activities~\cite{penzenstadlerSafetySecurityNow2014,chitchyan2016}.
A mapping study by Bambazek~\etal~\cite{bambazek2023RE} identified the Sustainability Awareness Framework (SusAF)~\cite{duboc2019we} as one of the most established methods.
It provides structured guidance for facilitating sustainability workshops and employs a set of guiding questions to help practitioners recognize possible sustainability impacts of software systems.

Early work on contextual RE introduced goal-oriented models~\cite{aliGoalbasedFrameworkContextual2010}.
The idea of modeling adaptation intents as sustainability goals was initially proposed by Gerostathopoulos~\etal~\cite{gerostathopoulos2022} and was subsequently used in a series of studies. 
For instance, $\mathcal{H}armonE$ leverages the approach to incorporate self-adaptive capabilities into MLOps pipelines to support long-term sustainability~\cite{bhattArchitectingSustainableMLOps2024,bhatt$$mathcalHarmonE$$Selfadaptive2026}.
Recent work has combined contextual goal models with proactive self-adaptation mechanisms, for example by integrating goal-oriented requirements with control-theoretic approaches to adapt system behavior under changing context~\cite{chenContextAwareProactiveSelfAdaptation2024}. They largely focus on qualities such as reliability or efficiency but do not explicitly address sustainability concerns.
Some studies discuss the notion of combining adaptive security and (partially) sustainability. For instance, Halabi~\etal~\cite{halabiAdaptiveCybersecurityGreen2022} propose to use adaptive cybersecurity for Green IoT focusing on the ``energy-efficient and environment-friendly paradigm''.
Hence, existing work almost exclusively focuses on the environmental dimension, treating sustainability, safety, and security as separate concerns. 


The  SAF Toolkit, developed by Lago~\etal~\cite{lagoSustainabilityAssessmentFramework2025}, provides comprehensive support for modeling sustainability and covers important system characteristics.
However, the SAF Toolkit lacks dedicated support for requirements, particularly for safety- and security-critical systems and self-adaptation. 
The solution we propose in this paper specifically addresses these challenges by extending the SAF Toolkit notation to explicitly capture security, safety, and health concerns, and by introducing novel mechanisms for handling and executing dynamic behavior based on security modes.

\ms{

}

\section{Approach}
\label{sec:approach}

To address the aforementioned problems and challenges, we extend the current DM notation with additional semantic elements (cf.~\citefig{our-dm-notation}).

It is important to note that the extended DMs intend to function in conjunction with established dependability engineering practices. 
We assume that rigorous threat modeling~\cite{xiongThreatModelingSystematic2019} (e.g., attack trees~\cite{gadyatskayaLimitedTechnicalBackground2025}) or hazard analysis~\cite{ericsonHazardAnalysisTechniques2005} (e.g., fault trees~\cite{veselyFaultTreeHandbook1981,ruijtersFaultTreeAnalysis2015}) has been performed, identifying vulnerabilities and safety concerns.

\begin{figure}
    \centering
    \includegraphics[width=0.78\linewidth]{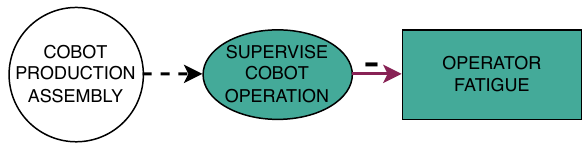}
    \caption{\textit{Individual} dimension in the running example.}
    \label{fig:indiv-run-example}
\end{figure}


\begin{figure*}[t!]
  \centering
  \begin{subfigure}[b]{0.49\textwidth}
    \captionsetup{font=footnotesize}
    \includegraphics[width=\textwidth]{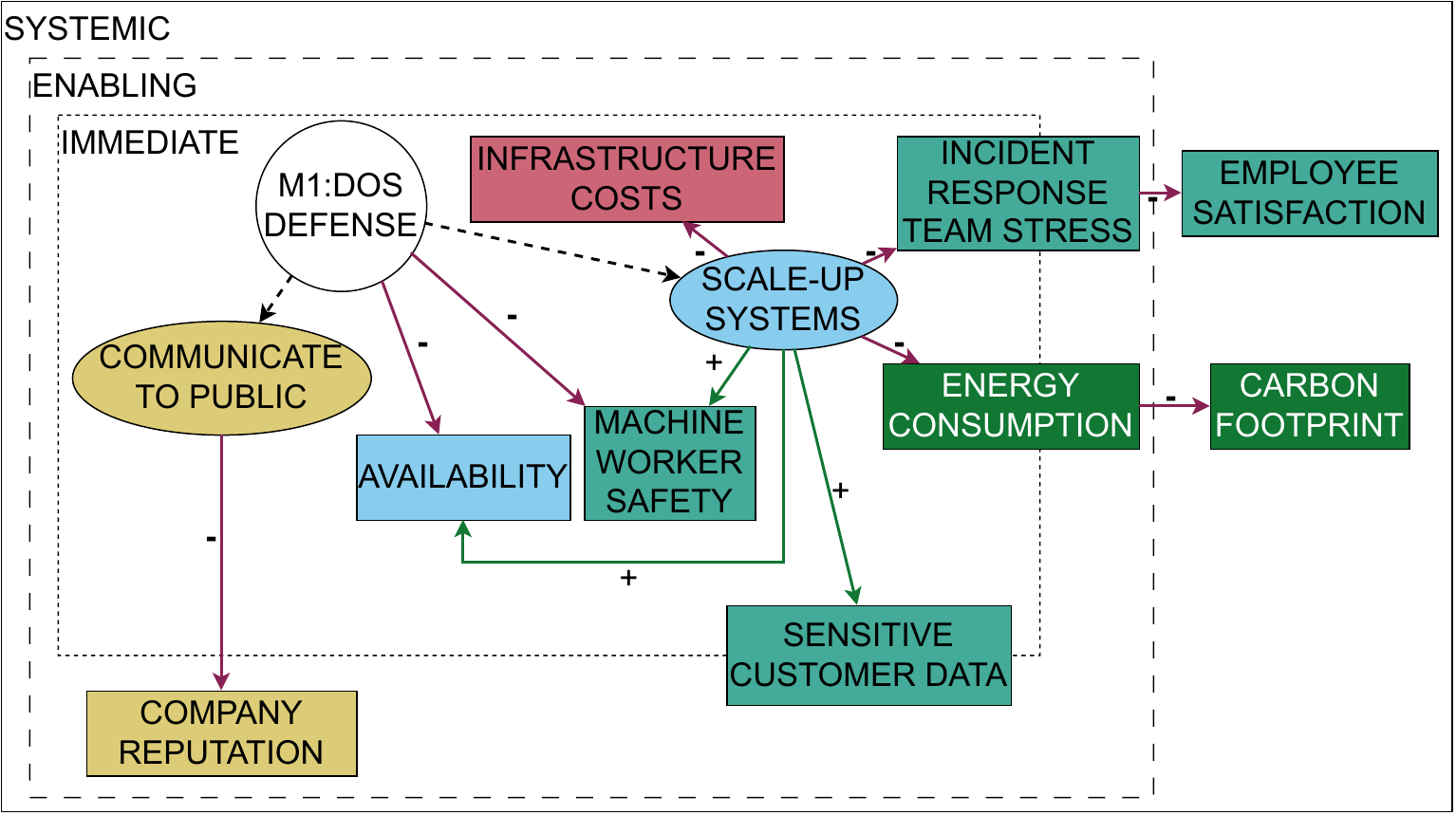}
    \caption{DM for ``DOS DEFENSE MODE''.}
    \label{fig:dos-mode}
  \end{subfigure}
  \hfill
  \begin{subfigure}[b]{0.4\textwidth}
    \captionsetup{font=footnotesize}
    \includegraphics[width=\textwidth]{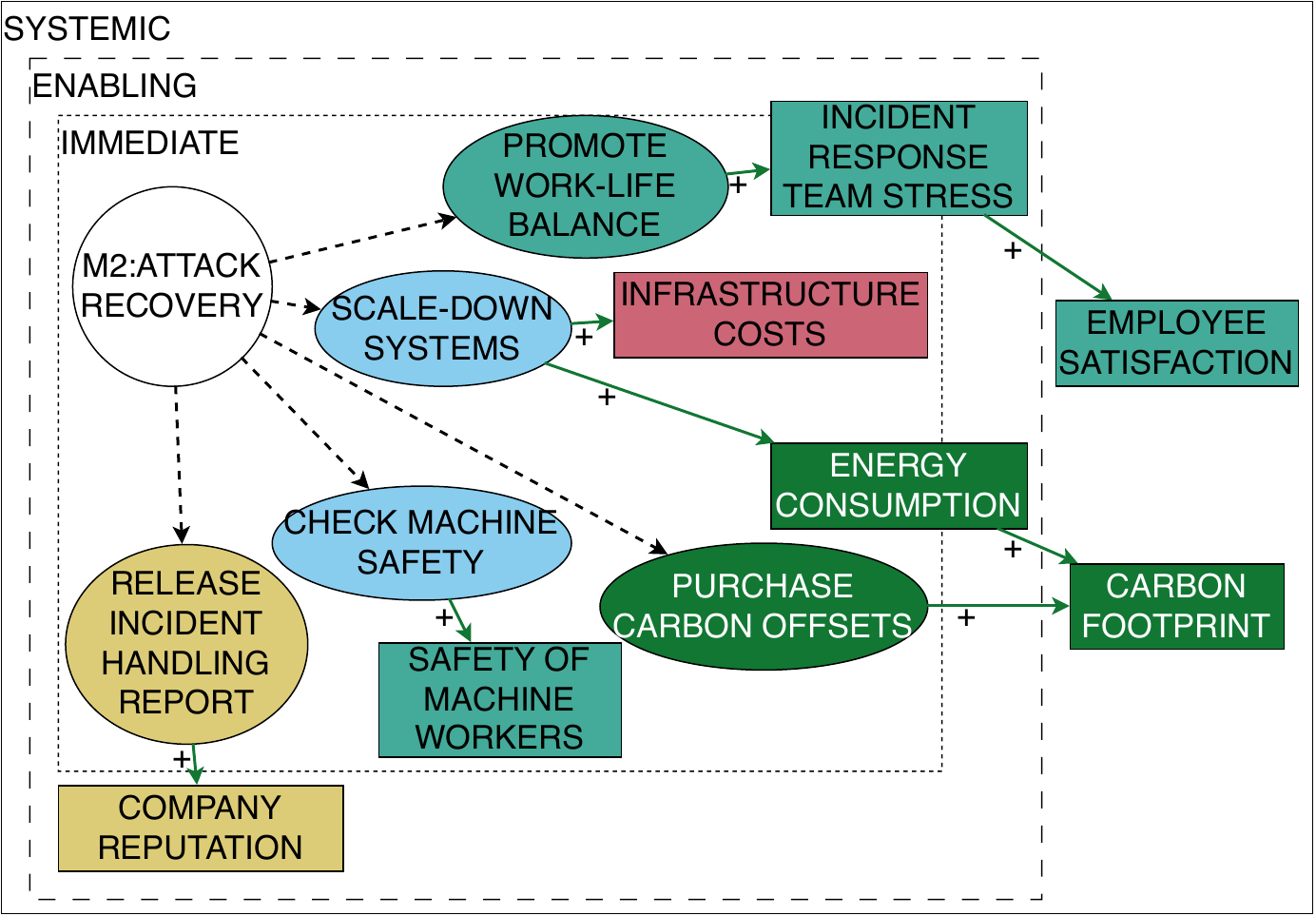}
    \caption{DM for ``Attack Recovery Mode''.}
    \label{fig:recovery-mode}
  \end{subfigure}
  \caption{DMs for the industrial use case.}
  \label{fig:main_figure}  
  \vspace{-0.1em}
\end{figure*}

\textbf{Extension 1 -- Individual Dimension:} The current notation for modeling sustainability goals using DMs consists of four dimensions: social, technical, environmental, and economic (cf.~\citesec{background}). 
While these four dimensions have proven to be sufficient for several use cases (e.g.,~\cite{bhatt$$mathcalHarmonE$$Selfadaptive2026}), we argue that from a safety perspective, it is beneficial to introduce a new fifth dimension: the \textit{individual}.
The \textit{individual} dimension is, among others, inspired by the SusAF~\cite{penzenstadlerSafetySecurityNow2014, dubocRequirementsEngineeringSustainability2020, betzSustainabilityAwarenessFramework2022}.
The dimension in SusAF covers the topics lifelong learning, privacy, agency, as well as health and safety~\cite{dubocRequirementsEngineeringSustainability2020}.
They define health as ``the state of a person's mental or physical condition''~\cite{betzSustainabilityAwarenessFramework2022}. 
For the health topic, the framework tries to answer the question of how the system can improve or worsen a person’s physical, mental, and/or emotional health.
The safety topic in SusAF refers to the protection from danger, risk, or injury~\cite{betzSustainabilityAwarenessFramework2022}.
More specifically, for the safety topic, it tries to answer how the system exposes (or protects) a person from physical harm and how it makes a person feel more (or less) exposed to harm. 
It also includes the question of what happens if the system is used in an unintended way and the influence of such an event on an individual's safety.
Hence, we suggest leveraging this already well-established and defined sustainability dimension in the (RE) sustainability community and combining it with the DM notation.
We propose to extend and incorporate the notion of the \textit{individual} dimension to capture an individual's safety and health more accurately over a long period of time.

The resulting notation is shown in~\citefig{our-dm-notation}. 
In particular, we propose to extend the \fwcomp{TYPES OF DIGITAL FEATURES OR REQUIREMENTS} with an additional \indcomp{FEATURE OR REQUIREMENT (INDIVIDUAL)} and add an additional component to the \fwcomp{TYPES OF DIGITAL FEATURES OR REQUIREMENTS} called an \indcomp{INDIVIDUAL CONCERN}.
With this additional dimension, we can model safety-relevant concerns in the DMs and address further requirements related to an individual's well-being.
Importantly, a single safety requirement can be met by different alternative features. 
The extended DM notation allows users to explicitly model these feature choices, making it easier to compare their unique sustainability and socio-technical trade-offs to select the most suitable option.
Applying this to our running example introduced in \citesec{background}, we can now model the machine worker's fatigue using our new notation elements (cf.~\citefig{indiv-run-example}).
We model the requirement \indcomp{SUPERVISE COBOT OPERATION} as a \indcomp{FEATURE OR REQUIREMENT (INDIVIDUAL)}. 
We explicitly link this requirement to the \indcomp{INDIVIDUAL CONCERN} \indcomp{OPERATOR FATIGUE} with a negative effect.
This visualizes that while the human's validation is necessary for safety (resolving false positives), the repetitive cognitive demand of this task negatively impacts the operator's long-term well-being and reliability.





\textbf{Extension 2 -- Security Modes:}
Security is a cross-cutting concern that intersects with all sustainability dimensions.
For example, user privacy (social),  sensor data privacy (environmental), resilience against cyberattacks (technical), and reduced risk of financial loss (economic).
Thus, security is not a standalone pillar, but rather an essential quality that influences and reinforces sustainable outcomes~\cite{penzenstadlerSafetySecurityNow2014}.
Security is not static, it is an evolving system property.
Our example of the Log4Shell vulnerability, and the DoS attack in our running example illustrate how requirements on a software system can change over the course of an attack.
The current DM notation's capability of modeling changing requirements (dynamic behavior) is limited~\cite{gerostathopoulos2022}:
``Throughout a project, a DM can be updated to reflect, e.g., changes in requirements or a better-informed understanding of the expected or actual effects.''
When modeling sustainable security, we can therefore already be certain that a model created with the current DM notation is subject to change.
Instead of assuming the DM model ``as is'' and updating it to changing requirements (suggested  in~\cite{gerostathopoulos2022}) once a security incident has already happened, we advocate for proactively leveraging the dynamics of an attack and modeling the software system from different security viewpoints already at design time.
Analogous to security threat modeling~\cite{xiongThreatModelingSystematic2019}, which involves analyzing potential threats at design time and then proposing appropriate mitigation techniques, we argue that by proactively investigating the timing dynamics of an attack already when modeling the adaptation intents, we can identify potential quality concerns and, most importantly, the corresponding changing system requirements beforehand.

We propose extending the DM notation by modeling security in the DMs using \textit{security modes}.
The \textit{security modes} contextualize a system based on different attack stages or attack surface (e.g., \fwcomp{vulnerable mode} vs \fwcomp{normal mode} for the web server in \citesec{background}). 
By using the \textit{security modes} during DM modeling, we actively explore the system space along the attack stages and capture the implications of each \textit{security mode} separately.
Each identified change is then modeled as part of the DM and does not need to be manually changed at runtime later on in the DM.

As a result, this yields multiple DMs, one per mode, for the same software system.
For the notation elements, we introduce \fwcomp{SOFTWARE ARCHITECTURE MODES} in \citefig{our-dm-notation} to support this extension.
For each mode, it is possible to capture the (security-relevant) concerns.
For the running example in \citesec{background}, because of this extension, it is possible to model the DMs individually for each security mode.
For each \textit{security mode}, we create a dedicated DM, thus capturing the implications during a DoS attack (\fwcomp{M1: DOS DEFENSE}), during attack recovery (\fwcomp{M2: ATTACK RECOVERY}), or during normal operation times (\fwcomp{M3: NORMAL MODE}).
While we focus on security incidents here, this extension natively supports safety contexts (e.g., an ``Emergency Shutdown Mode'').



\textbf{Compensatory Adaptation:}
With our extensions (especially the \textit{security modes}), we provide conceptually a new perspective on the evolution of quality concerns and the adaptations over time.
Related work expressed the sustainability boundaries (cf. \fwcomp{Max} and \fwcomp{Min} in \citefig{our-intent}) as the target adaptation space for quality concerns.
This means that adaptations and effects should only happen within the sustainability boundaries to meet the initial adaptation intent.
We argue in this paper that this is not realistic in safety and security scenarios, where priorities shift over time.

We illustrate this using the cobots running example (cf.~\citesec{background}): 
In the event of a safety emergency (e.g., confirmed alert with the cobot), the priorities of an organization/system shift to protect human lives at all cost, as this is of utmost importance at this very moment.
During this shift in priorities, other sustainability quality concerns degrade.
For instance, there might be a direct economic loss because of the immediate shut down of the cobot manufacturing line.
One can observe a similar shift in priorities in the event of a security incident.
The average impact of a successful attack on a company’s ICS costs a company \$5 million, 50 days of downtime, and it takes around 191 days for an organization to fully recover from an incident~\cite{alladiIndustrialControlSystems2020}.
Hence, for an organization to stay economically competitive, it will prioritize warding off any attack that compromises the availability of the production line (e.g., a DoS attack) and trade it off with other quality concerns.
For instance, the employee satisfaction of a security team will degrade because of overtime to close the vulnerability.
Hence, we see the shift of priorities as unavoidable and as a more realistic viewpoint in these types of systems.




By modeling the target system over time at design time, we can identify the aforementioned trade-offs in quality concerns preemptively and deal with the degradation of quality concerns in extreme situations.
Consider \citefig{our-intent}: We can use the security modes to anticipate potential degradations at design time. 
For instance, during the \fwcomp{DOS DEFENSE MODE}, we then know that certain quality concerns are not within our sustainability boundaries.
We can use this knowledge to introduce and enter a \fwcomp{recovery mode} where we make up for the underperforming time, i.e., we enter a compensation stage, to make up for the underperforming (but necessary) times.




%



\section{Evaluation}
\label{sec:eval}

\subsection{Methodology}

To validate our proposed approach, we use the extended DM notation in a series of use cases.
Due to space restrictions, we report only one use case: an industrial proof-of-concept validation. 
The industrial proof-of-concept use case is motivated by our industry partner (details in context description). 
Using the extended DM notation in a realistic use case scenario showcases the practical relevance of our approach.

The SASs research community has accumulated several ``exemplars'' over the years~\cite{vogelSoftwareEngineeringSelfAdaptive2026}.
The idea is to use SASs exemplars to promote active research, share common adaptation problems, and facilitate research in general~\cite{shinPlatooningLEGOsOpen2021}.
In our evaluation (due to space restrictions not reported here), we tested our approach also using the  DeltaIoT~\cite{iftikharDeltaIoTSelfAdaptiveInternet2017} and SWIM~\cite{morenoSWIMExemplarEvaluation2018} exemplars.

Our approach outlines two major extensions (\textit{Extension 1 -- Individual Dimension} \& \textit{Extension 2 -- Security Modes}) and an impact on the evolution of adaptations (\textit{Compensatory Adaptation}). 
To answer \textbf{RQ2} (feasibility), we report for one use case the implications of using the two extensions and what the new perspective for the specific use case implies.
This allows for a well-informed decision on whether our proposed extended DM notation is indeed feasible.

\subsection{Industrial Use Case}
\label{sec:industrial-example}



The use case is motivated by our industry partner. \indpart is a large machine manufacturing company, operating in over 80 countries, with several thousand employees, and is one of the leading manufacturers of \machines. 
In practice, these machines operate largely isolated from public networks (e.g., in segmented or even air-gapped production environments) and are protected by established security measures, so a scenario such as the one modeled below is highly unlikely to occur in the field.
Nevertheless, because downtime (e.g., due to a security incident) could severely disrupt operations and cause significant financial losses, and because the safety of workers operating close to the machines is paramount, such worst-case scenarios are exactly what practitioners must reason about at design time. 
Therefore, the company serves as an excellent case study.
We use our extended notation to model an injection molding machine under a DoS attack (e.g., from a peripheral network). 
\citefig{dos-mode} shows the sustainability goals during the DoS attack, and \citefig{recovery-mode} shows them during recovery from the attack.

\textit{Extension 1 -- Individual dimension:}
During the attack, the infrastructure will be scaled up.
Although this requirement has a positive effect on \uppercase{\fwcomp{availability}}, \indcomp{machine worker safety}, and the security of \indcomp{sensitive customer data}, it also has several negative side effects.
As infrastructure costs rise, the \indcomp{incident response team stress} increases, and the \uppercase{\fwcomp{energy consumption}} rises.
Furthermore, disclosing the attack to the public might influence the company's reputation.
However, in the interest of safety and availability, all these negative effects are unavoidable short-term effects.
During recovery, we are able to capture (among other concerns) the inverse safety intents of the DoS attack: the company can \indcomp{promote work-life balance} to eventually reduce \indcomp{incident response team stress} and increase \indcomp{employee satisfaction}.

\textit{Extension 2 -- Security modes:}
In the given scenario, the system is modeled to operate in two security- and safety-relevant operational modes: 
\fwcomp{M1:~DOS DEFENSE} -- i.e., during the DoS attack; \fwcomp{M2: ATTACK RECOVERY} -- i.e., after the attack has been warded off. 
As a result, we are able to capture the timing-specific implications of the different modes.

\textit{Compensatory Adaptation:}
Based on an analysis of the previously outlined defense mode, it is possible to address neglected sustainability quality concerns to restore balance in the adaptation intent.
When the system is in defense mode (\fwcomp{M1: DOS DEFENSE}), the priority is to minimize the impact of the attack; hence, the immediate actions taken by the incident response team and the SAS are scaling up the infrastructure.
In particular, the negative effects in \fwcomp{M1} (cf. red arrows in~\citefig{dos-mode}) should be used to identify potential countermeasures (i.e., positive effects) to meet the overall sustainable adaptation intent. 
For instance, to recover from the stress of the DoS attack, the incident response team can take additional time off to make up for the distress.
\section{Discussion}
\label{sec:disc}



The goal of our work is to find novel ways for modeling adaptation intents as sustainability goals for safety- and security-critical scenarios.
The implications of our evaluation, applying the extended notation to the industrial use case, can be summarized as follows: \vspace{0.2em}

With the newly added individual dimension, we were able to capture and model not only safety concerns (e.g.,~\fwcomp{MACHINE WORKER SAFETY} in \citefig{dos-mode}), but we can also represent a variety of human-centric concerns, such as the \fwcomp{INCIDENT RESPONSE TEAM STRESS}.
Without our DM notation, these aspects were either not captured or only implicitly and coarsely captured through other dimensions.
Our extension with the individual dimension directly supports sustainable safety adaptations in SASs.\vspace{0.2em}



Similar to threat modeling~\cite{xiongThreatModelingSystematic2019}, explicitly analyzing worst-case scenarios (cf. defense mode in~\citesec{eval}) systematically exposes attack impacts and provides the structural basis to design compensatory adaptation mechanisms (cf. recovery modes in~\citesec{eval}). 
Consequently, our approach identifies (negative) effects and proactively finds ways to adapt in accordance with sustainability goal intents.
A visualization of the quality of a sustainability goal over time with active compensation is shown in \citefig{our-intent}. 
In a realistic scenario, a cyber attack will always shift priorities, thereby resulting in underperforming quality for some sustainability goals (red area).
However, our approach allows us to identify those negative effects and, at the same time, plan for adaptation opportunities for compensation (green area).\vspace{0.2em}

Taking all of the above into consideration, the answer to \textbf{RQ1} can be summarized as follows:
\begin{tcolorbox}[myrqbox]
     \textbf{Answer to RQ1} -- Two extensions are needed to express the safety- and security-aware adaptation intent: First, an additional \textit{individual} dimension to model safety and human-centric concerns. Second, \textit{security modes} enable us to express shifting sustainability concerns over time.
\end{tcolorbox}

Building on our conceptual extensions identified in \textbf{RQ1}, \textbf{RQ2} examines how the extended DM notation behaves when applied in practice. 
Particularly, whether the new individual dimension and the use of security modes are usable in practice.
We validated the feasibility through application in an industrial domain. 
The successful modeling leads to the following conclusion for \textbf{RQ2}:

\begin{tcolorbox}[myrqbox]
    \textbf{Answer to RQ2} -- The extended DM notation proved feasible in an industrial manufacturing scenario.
    The newly introduced individual dimension explicitly captures human-centric safety and health concerns, such as operator fatigue and incident response team stress.
    Security modes allow modeling multiple operational modes, such as normal operation, different attack phases, and recovery.
    By capturing system behavior over time, the DMs make security dynamics explicit, and reveal mode-specific trade-offs and compensation strategies.
\end{tcolorbox}
\vspace{0.2em}

\section{Conclusion}
\label{sec:conclusion}

In this paper, we present an extended DM modeling notation to support SAS safety and security scenarios.
The extension supports modeling the adaptation intent using sustainability goals for security and safety scenarios by adding an additional \textit{individual} dimension and leveraging \textit{security modes} in the DMs.
Our evaluation provides a first indication of the feasibility of our approach.
However, extending the current notation is only a first step towards achieving a comprehensive, sustainable security and safety support.
We envision a full integration into the SAF Toolkit pipeline~\cite{lagoSustainabilityAssessmentFramework2025} in the future to facilitate true sustainable security and safety adaptations.



\clearpage
\newpage
\balance
\bibliographystyle{IEEEtran}
\bibliography{main}

\end{document}